\documentclass[a4paper,11pt]{article}
\pdfoutput=1 

\usepackage{jcappub} 

\usepackage[T1]{fontenc} 

\renewcommand{\a}{\alpha}
\renewcommand{\b}{\beta}
\renewcommand{\d}{\delta}
\renewcommand{\gg}{\gamma}

\renewcommand{\ss}{\sigma}

\newcommand{\be}{\begin{equation}}
\newcommand{\ee}{\end{equation}} 
\newcommand{\eei}{\end{equation}\indent\indent}
\newcommand{\bc}{\begin{center}}
\newcommand{\ec}{\end{center}}
\newcommand{\ber}{\begin{eqnarray}}
\newcommand{\ear}{\end{eqnarray}}
\newcommand{\ba}{\begin{array}}
\newcommand{\ea}{\end{array}}

\def\case#1/#2{\textstyle\frac{#1}{#2} }

\title{\boldmath Cosmological dynamics: from the Eulerian to the Lagrangian frame -- \\ I. Newtonian approximation}

\author[a,1]{Eleonora Villa,\note{Corresponding author.}}
\author[b,c]{Sabino Matarrese,}
\author[a]{and Davide Maino}

\affiliation[a]{Dipartimento di Fisica, Universit\`a di Milano,\\via Celoria 16, 20154 Milano, Italy}
\affiliation[b]{Dipartimento di Fisica e Astronomia "G. Galilei", Universit\`a degli Studi di Padova and INFN Sezione di Padova,\\via Marzolo 8, 35131 Padova, Italy}
\affiliation[c]{Gran Sasso Science Institute, INFN,\\ viale F. Crispi 7, 67100 L'Aquila, Italy}
\emailAdd{eleonora.villa@unimi.it}
\emailAdd{sabino.matarrese@pd.infn.it}
\emailAdd{davide.maino@mi.infn.it}

\abstract{We analyse the non-linear gravitational dynamics of a pressure-less fluid in the Newtonian limit of General Relativity 
in both the Eulerian and Lagrangian pictures. Starting from the Newtonian metric in the Poisson gauge, we transform to the synchronous 
and comoving gauge and obtain the Lagrangian metric within the Newtonian approximation. Our approach is fully non-perturbative, 
which implies that if our quantities are expanded according to the rules of standard perturbation theory, 
all terms are exactly recovered at any order in perturbation theory, only provided they are Newtonian. 
We explicitly show this result up to second order and in both gauges. Our transformation clarifies the meaning of the change of spatial 
and time coordinates from the Eulerian to the Lagrangian frame in the Newtonian approximation.}

\begin{document}
\maketitle
\flushbottom

\section{Introduction}
It is believed that the distribution of matter in the early Universe was very smooth, the best indication being the tiny fluctuations in the Cosmic Microwave Background. 
However, the distribution of matter in the Universe at present time is inhomogeneous on scales below about 100 $h^{-1}$ Mpc ($h$ being the Hubble constant 
in units of 100 km s$^{-1}$ Mpc$^{-1}$) and the gravitational dynamics is non-linear below 
about 10 $h^{-1}$ Mpc. The theoretical study of structure formation connects the early Universe with that observed today. 
The gravitational instability is governed by the equations provided by General Relativity (GR) but, for most purposes, we can use the Newtonian approximation, 
namely the weak-field and slow-motion limit of GR. These conditions are verified at small scales, well inside the Hubble radius, where the peculiar gravitational 
potential $\varphi_g$, divided by the square of the speed of light to obtain a dimensionless quantity, remains much less than unity, while the peculiar velocity never 
become relativistic. The peculiar gravitational potential is related to the matter-density fluctuation $\delta$ via the cosmological Poisson equation
\begin{equation}
\nabla^2\varphi_g=4\pi G a^2 \rho_b \delta
\end{equation}
where $\rho_b$ is the Freidmann-Robertson-Walker (FRW) background matter density, $\delta=\left(\rho-\rho_b\right)/\rho_b$ is the density contrast and $a(t)$ is the 
scale-factor, which obeys the Friedmann equation. This implies that for a fluctuation of proper scale $\lambda$,
\begin{equation}
\frac{\varphi_g}{c^2} \sim \delta\left(\frac{\lambda}{r_H}\right)^2 \;,
\end{equation}
where $r_H=cH^{-1}$ is the Hubble radius. This very fact tells us that the weak-field approximation does not necessarily imply small density fluctuations, 
rather it depends on the ratio of the perturbation scale $\lambda$ to the Hubble radius.
That is why Newtonian gravity is widely used to study structure formation at small scales, also in the non-linear regime. The evolution of perturbations is dealt 
with analytically, within perturbation theory, or numerically, by means of  N-body simulations. 

However, there are situations, even within the Hubble horizon, where the Newtonian treatment is not well suited. The Newtonian approximation of 
GR, which consists in perturbing the time-time component of the space-time metric by an amount $2\varphi_g/c^2$ fails to produce an accurate 
description of photon trajectories: it is well know that the Newtonian estimate of the Rees-Sciama effect and of gravitational lensing is incorrect by 
a factor of 2. The correct calculation involves the weak-field limit of GR, which is valid for slow motions of the sources of the gravitational field, but allows 
test particles to be relativistic. The related metric is perturbed also in the space-space component of the space-time metric by an amount $-2\varphi_g/c^2$. 

Above all, the upcoming galaxy surveys will probe increasingly large scales, approaching the Hubble radius, where the Newtonian approximation 
is no longer valid and a relativistic treatment is therefore mandatory, both to study the grow of structures and to analyse cosmological observations at these scales. 
The study of gravitational instability in the non-linear regime and in GR is of course very challenging and it cannot be accomplished without some approximations. 
The most widely used is cosmological relativistic perturbation theory, truncated at first or at second-order. In the perturbative framework, the second-order approach 
in calculations is crucial: the linear modes generate non-linear ones, many of which are not taken into account in Newtonian theory, and play an important role in the 
dynamics of perturbations as well as in the calculations of the gravitational lensing and redshift space distortions, see ref.~\cite{veneziano1}-\cite{umeh} and refs. therein. 

On the other hand, a description of the fully non-linear dynamics of the perturbations can be carried out by means of effective fluid description for small-scale non-linearities 
in the framework of GR, as proposed in ref.~\cite{baumann} or by means of an effective field theory approach, as proposed in ref.~\cite{porto} using a Lagrangian approach. 
Also, exact analytical solutions of Einstein's equations, can be important tools to investigate non-linear effects, e.g. in the path of photons throughout inhomogeneities, 
although in simplified contexts or in presence of some symmetry in the problem, see e.g. refs.~\cite{meures1} and~\cite{meures2}.

Recently, a GR analysis of galaxy clustering, refs.~\cite{BMR2005}-\cite{raccanelli} and refs. therein, has drawn attention to other relativistic effects in the observations, 
such as the gauge dependence of the galaxy bias, which become sizeable near the Hubble horizon, see ref.~\cite{brunibias} and ref.~\cite{wands}. In this respect, a 
second-order calculation actually shows a gauge-dependence in the matter density also in the Newtonian expressions, see ref.~\cite{BMPR2010} and ref.~\cite{colombi92}.

The connection between Newtonian N-body simulations and GR is also of great interest today: in ref.~\cite{chisari} the authors provide a dictionary for how to interpret 
the outputs of numerical simulation, run using Newtonian dynamics, with respect to GR at first order. In ref.~\cite{durrerNbody} a formalism for GR N-body simulations which go beyond standard perturbation theory is provided in the weak-field limit of GR, which is then used in ref.~\cite{durrerdlz} to compute the distance-redshift relation for a plane-symmetric universe. For the interpretation of N-Body simulations and the connection to the Lagrangian and Eulerian perspectives see ref.~\cite{RigopoulosNbody}, where the relativistic corrections to Newtonian cosmology are analysed with the gradient expansion method on scales close to the Hubble scale.\\
A fully second-order implementations of GR corrections to N-body simulations is still missing and would certainly be useful, although second-order perturbation theory 
should not to be considered as an exhaustive study for non-linear dynamics.

Rather, we should seek for a relativistic and non-perturbative approach, capable to disentangle the Newtonian from the GR contributions. 
An alternative approximation scheme is well-suited for this purpose: the Newtonian analysis can be improved with the post-Newtonian (PN) approximation of GR, 
which provides the first relativistic corrections for a system of slowly moving particles bound together by gravitational forces and thus it can be used to account for the moderately non-linear gravitational field generated during the highly non-linear stage of the evolution of matter fluctuations on intermediate scales. It is a crucial improvement of both the aforementioned approximations, as it could bridge the gap between relativistic perturbation theory and Newtonian structure 
formation, providing a unified approximation scheme able to describe the evolution of cosmic inhomogeneities from the largest observable scales to small ones, 
including also the intermediate range, where the relativistic effects cannot be ignored and non-linearity starts to be relevant. Nonetheless, few attempts have been 
made so far to go beyond the Newtonian approximation on non-linear scales. A relevant difficulty of this scheme is that in the PN framework the background is not 
merely the FRW metric but a Newtonian metric yet describing non-linearities. 
This very fact and has so far prevented from proceeding in this direction because of its computational complexity, except for symmetric situations. In ref.~\cite{VMM} 
the PN solution of the Einstein field equations describing the non-linear cosmological dynamics of plane-parallel perturbations in the synchronous and comoving 
gauge was obtained. It extends the Zel'dovich approximation, which, in turn, is exact for non-linear plane-parallel dynamics in Newtonian gravity.
The PN approximation has by construction a direct correspondence with Newtonian quantities: the PN expressions are sourced only by the non-linear Newtonian 
terms which can be extracted e.g. from N-body simulations (or by approximate analytical expressions obtained via the Zel'dovich  approximation). For a recent 
attempt to include PN-type corrections from N-body simulations see ref.~\cite{brunidrag} and ref.~\cite{brunilensing}.

In this paper we consider the non-linear dynamics of cosmological perturbations of an irrotational collisionless fluid, the FRW background being the Einstein-de 
Sitter model. We discuss the connection between GR and Newtonian gravity in the Eulerian and in the Lagrangian picture. Then we provide the transformation rule 
between the Eulerian and the Lagrangian frames: it is fully four-dimensional and non-perturbative and clarifies the role of the transformation of the time and spatial 
coordinates. Our approach here is different from the standard perturbative gauge transformation in GR, see refs.~\cite{MMB}, \cite{MBgeo} and \cite{bardeen}, and from the spatial coordinate transformation of Newtonian theory.

Most derivations of the Newtonian limit of GR are coordinate-dependent, thus a precise understanding of the Newtonian correspondence between the Eulerian 
and the Lagrangian frame has to be considered as the starting point e.g. for studying the gauge dependence 
when we want to add GR corrections in a perturbed space-time from a non-perturbative perspective. In a forthcoming paper we will extend this analysis to the PN approximation.

This paper is organised as follows: in the next two sections we review the study of the non-linear dynamics in the Newtonian limit of GR in the Eulerian and 
Lagrangian pictures, respectively. In section 4 we present the coordinate transformation and in section 5 the second-order expansion of our quantities in 
perturbation theory. We first review the Newtonian results in the Eulerian picture. Then we show that our calculation fully recovers the solution for the 
Lagrangian metric in the Newtonian limit. Conclusions are drawn in the last section.

Indices notation: we use $A, B, ...$ for spatial Eulerian indices, $\a, \b, ...$ for spatial Lagrangian and $a, b, ...$ to indicate space-time indices in any gauge. 
The superscript or subscript $\mathcal{E}$ and $\mathcal{L}$ stands for Eulerian and Lagrangian, respectively, when needed for clarity.

\section{The Newtonian approximation in the Eulerian picture}
\label{euler}
 
In the Eulerian picture the matter dynamics is described with respect to a system of coordinates not comoving with the the matter. 
In a uniformly expanding Universe, all physical separations scale in proportion with a cosmic scale factor $a(t)$. Even though 
the expansion is not perfectly uniform, it is perfectly reasonable to factor out the Hubble expansion and we do this by using FRW 
comoving spatial coordinates and conformal time defined as
\begin{equation}
\begin{split}
x =\frac{r}{a(t)}\,,
\qquad
d\eta= \frac{dt}{a(t)}  \,,
\end{split}
\end{equation}
where $r$ is proper spatial coordinates, $t$ is the cosmic time and $a(t)$ the scale-factor.
With this choice all physical quantities appearing in the equations are measured by observers comoving with the FRW background expansion. The coordinate velocity is 
\begin{equation}
v^A=\frac{dx^A}{d\eta}=\frac{dr^A}{dt}-H r^A\,,
\end{equation}
where $H=\partial_t a/a$ is the Hubble parameter. It is the physical velocity of the matter minus the Hubble expansion, i.e. the peculiar velocity.

The Newtonian equations in the Eulerian picture read
\begin{align} 
0&=\frac{\partial\delta_\mathcal{E}}{\partial\eta}+v^C\partial_C\delta_\mathcal{E}+\partial_Av^A(1+\delta_\mathcal{E}) \label{newtcont}\\
0&= \frac{\partial v_{K}}{\partial \eta}+v^C\partial_C v_K+\mathcal{H}v_{K} +\partial_K\varphi^\mathcal{E}_g \label{newteuleq}\\
\nabla^2\varphi^\mathcal{E}_g &=4\pi Ga^2\rho_b\delta_\mathcal{E} \label{newtpoiss}\,,
\end{align}
where $\mathcal{H}= \partial_\eta a /a$. For the irrotational dust considered here we have also the condition
\begin{equation}
\epsilon^{ABC}\partial_Av_B=0\,,
\end{equation}
where $\epsilon^{ABC}$ is the totally antisymmetric Levi-Civita tensor relative to the Euclidean spatial metric, such that $\epsilon^{ABC}=1$, etc...

The fundamental variables in the Eulerian picture are the velocity and density field, evaluated at the Eulerian coordinates $x^A$.
The same equations can be obtained in the GR framework: the continuity and Euler equations are the lowst-order equations 
in the $1/c^2$ expansion of $\nabla^a T _{a0}$ and $\nabla^a T_{aK}$ respectively. The time-time component of the Einstein equations
\begin{equation} 
 R^0_0=\frac{8\pi G}{c^4}\left(T^0_0-\frac{1}{2}T\right)
\end{equation}
reduces to the Poisson equation and implies that the perturbation in the time-time component of the metric tensor coincides with the Newtonian gravitational potential.

The line-element in the Newtonian approximation of GR in the Eulerian picture then reads
 \begin{equation}\label{newtds^2E}
ds^2=a^2\left[-\left(1+2\frac{\varphi_g^\mathcal{E}}{c^2}\right)c^2 d\eta^2 +\delta_{AB}dx^A dx^B\right].
\end{equation}
The point of view illustrated above is \textit{purely Newtonian}: the matter is viewed as responsive to the gravitational field given by $\varphi_g$ 
and the Newtonian order in the metric is established considering just the equations of motion for the fluid, eqs. \eqref{newteuleq} and 
\eqref{newtcont}, where only $\gamma^\mathcal{E}_{00}=-\left(1+2\varphi_g^\mathcal{E}/c^2\right)$ is required, with $\varphi_g^\mathcal{E}$ 
satisfying the Poisson equation \eqref{newtpoiss}. The line element in eq. \eqref{newtds^2E} is our lowest, i.e. Newtonian, order approximation 
in the Poisson gauge as defined in ref.~\cite{bert96}\footnote{We remark that the line element in eq. \eqref{newtds^2E} is not referred to the so-called 
longitudinal gauge, where vector and tensor modes are set to zero by hand at all orders and only the scalar mode in the spatial metric is present. 
Actually, it is not even a gauge, since only one among the six physical degrees of freedom in the metric are allowed.}. Accordingly, in our approach 
the PN line element in the Poisson gauge contains a divergenceless vector contribution in the space-time component and a scalar 
contribution in space-space component of the metric:
\begin{equation} \label{PNds^2E}
ds^2=a^2\left[-\left(1+\frac{2\varphi_g^\mathcal{E}}{c^2}+\frac{\Phi^\mathcal{E}}{c^4}\right) c^2 d\eta^2 +\frac{\omega^\mathcal{E}_A}{c^3}c 
d\eta dx^A+\left(1+\frac{2\varphi_g^\mathcal{E}}{c^2}\right)\delta_{AB} dx^Adx^B\right].
\end{equation}
The PN terms here are given by the lowest-order of the $1/c^2$ expansion of all the Einstein equations. In this gauge the transverse and 
traceless tensor modes, generated by the non-linear growth of scalar perturbations, appear at PPN level ($\mathcal{O}\left(1/c^4\right)$) only. 

For a PN calculation in the Eulerian picture see refs.~\cite{CM2005} and~\cite{kof}. We postpone a more detail discussion about the PN 
approximation in this gauge to our subsequent paper, ref.~\cite{VMMPN}.

\section{Newtonian approximation in the Lagrangian picture}
\label{lagr}
In the Lagrangian picture the dynamics is described with respect to a synchronous and comoving system of coordinates attached to the matter: 
at some arbitrarily chosen initial time we label the fluid elements by spatial coordinates $q^\a$; at all later times, the same element is labelled by 
the same coordinate value. The time coordinate is the proper time $\tau$ along the fluid trajectories. This frame is related to the set of observers 
comoving with the fluid elements: their proper time along the trajectories and their fixed initial labels are taken to define the coordinate values.

The spatial Lagrangian position vector of the fluid elements is given by the curve $q^\a$ = const., implying that the velocity of matter vanishes in 
Lagrangian coordinates. On the other hand, the Eulerian, time-dependent position vector could be expressed in terms of Lagrangian spatial coordinates as
\begin{equation}\label{newtmap}
\mathbf{x}(\mathbf{q},\tau)=\mathbf{q}+\mathbf{\mathcal{S}}(\mathbf{q},\tau)
\end{equation}
and the peculiar velocity is
\begin{equation}
\mathbf{v}=\frac{d\mathbf{\mathcal{S}}}{d\tau}\,.
\end{equation}
Similarly to the Eulerian trajectory, also the peculiar velocity of the matter $d\mathbf{x}/d\tau$, can be expressed in terms of the initial labels of the 
fluid element $q^\a$ through the mapping in eq. \eqref{newtmap}. 
Note that the relation above refers to an inhomogeneous universe and is fully non-perturbative. For a perfectly uniform expansion, the comoving 
position vector $\mathbf{x}$ is fixed in time and coincides with its initial, i.e. Lagrangian, coordinate value $\mathbf{q}$. On the contrary, in a perturbed 
universe it changes with time as irregularities grow, in the way described by the relation above. 
Therefore, all the information about the evolution of the perturbations is contained in the mapping relation \eqref{newtmap}: one can equivalently write the 
equations of motion of the fluid in terms of either the displacement vector $\mathbf{\mathcal{S}}$, as in ref.~\cite{catelan95L}, or in terms of the Jacobian matrix of the map
\begin{equation}\label{jacomap}
\mathcal{J}^A_\a=\frac{\partial x^A}{\partial q^\a}=\delta^A_\a+\frac{\partial S^A}{\partial q^\a}\,,
\end{equation}
as in ref.~\cite{mater}, where $\mathcal{S}^A_\a=\partial S^A/\partial q^\a$ is called the deformation tensor.

In GR, for a pressureless fluid, the time coordinate $\tau$ can be defined in such a way that it satisfies the following two conditions: any 
hypersurface $\tau$ = const. is orthogonal to the world-line of the fluid elements in any point and the variation of $\tau$ along each world-line 
coincide with the proper time variation along it. In the synchronous and comoving gauge, comoving hypersufaces are orthogonal to the matter 
flow ($\gamma^\mathcal{L}_{0\b}=0$) and coincide with the synchronous ones,  ($\gamma^\mathcal{L}_{00}=-1$), orthogonal to geodesics. 
The line element then reads
\begin{equation}
ds^2=a^2\left[-c^2 d\tau^2 +\gamma^\mathcal{L}_{\a\b}dq^\a dq^\b\right].
\end{equation}

Let us start by defining the peculiar velocity-gradient tensor $\vartheta^\alpha_{~\beta}$, given by
\begin{equation}\label{thetaL}
\vartheta^\a_\b=a u^\a_{;\b}-\mathcal{H}\delta^\a_\b=\frac{1}{2}\gamma^{\sigma\a}\gamma^{'}_{\sigma\b}\,,
\end{equation}
where the semicolon indicates the covariant derivative and the last equality holds in this gauge only.  

The Einstein equations in this gauge can be written as the energy constraint
\be \label{energy}
\vartheta^2 - \vartheta^\mu_{~\nu} \vartheta^\nu_{~\mu} + 4 \mathcal{H}
\vartheta -16 \pi G a^2\rho_b \delta  =  - c^2 \,^{(3)}\!{\cal R} 
\,,
\ee
the momentum constraint
\be \label{mom}
\mathcal{D}_\a\vartheta^\alpha_{~\beta} = \partial_{\beta}\vartheta \,, 
\ee
and the evolution equation
\begin{equation} \label{evol}
 \vartheta^{\a'}_{\b}+2\mathcal{H}\vartheta^{\a}_{\b}+\vartheta\vartheta^{\a}_{\b}+\mathcal{H}\vartheta\d^{\a}_{\b}-\frac{3}{2}\mathcal{H}^2\d\d^{\a}_{\b} 
 = -c^{2}\,^{(3)}\!\mathcal{R}^{\a}_{\b}\,.
\end{equation} 
Here $^{(3)}\!{\cal R}^\alpha_{\beta}$ is the conformal Ricci tensor of the three-dimensional space, $\mathcal{D}_\a$ is the covariant derivative corresponding 
to the metric $\gamma^\mathcal{L}_{\alpha\beta}$ and primes denote differentiation with respect to the coordinate time 
$\tau$. 
After replacing the density from the energy constraint, the evolution equation can be rewritten as
\begin{equation} \label{evolmater}
 \vartheta^{\a'}_{\b}+2\mathcal{H}\vartheta^{\a}_{\b}+\vartheta\vartheta^{\a}_{\b}+\frac{1}{4}\left(\vartheta^{\mu}_{\nu}\vartheta^{\nu}_{\mu} - 
 \vartheta^{2}\right)\d^{\a}_{\b}= -\frac{c^{2}}{4}\left[4\,^{(3)}\!\mathcal{R}^{\a}_{\b}-\,^{(3)}\! \mathcal{R}\d^{\a}_{\b}\right].
\end{equation}
The trace part of the evolution equation combined with the energy constraint to eliminate $^{(3)}\!{\cal R}$ gives the Raychaudhuri equation, which describes 
the evolution of the peculiar volume expansion scalar and reads
\begin{equation} \label{ray}
 \vartheta'+\mathcal{H}\vartheta+\frac{1}{3}\vartheta^2+ \sigma^\lambda_{~\rho}\sigma_\lambda^{~\rho} +4 \pi G a^2 
\rho_b \delta=0\,,
\end{equation} 
where $\sigma^\alpha_{~\beta} \equiv \vartheta^\alpha_{~\beta} - {1 \over 3} \delta^\alpha_{~\beta} \vartheta$ is the shear tensor, i.e. 
the trace-free part of the velocity-gradient tensor.

Finally, mass conservation implies 
\be 
\delta ' + \left(1 + \delta \right)\vartheta = 0 \;, 
\ee
which in this gauge can be solved exactly, by virtue of eq. \eqref{thetaL}. The solution is
\be \label{deltasolution}
\delta({\bf q}, \tau) = (1 + \delta_{in}({\bf q})) \left[\gamma({\bf q}, \tau)/ 
\gamma_{in} ({\bf q}) \right]^{-1/2} - 1 \,,
\ee
where $\gamma$ is the determinant of the metric $\gamma^\mathcal{L}_{\alpha\beta}$. The main advantage of this formalism, 
adopted in ref.~\cite{mater}, is that there is only one dimensionless variable in the equations, namely the spatial metric tensor $\gg^\mathcal{L}_{\a\b}$, 
and therefore there can be no extra powers of $c$ hidden in the definition of different quantities.

The Newtonian approximation is obtained in the $c\rightarrow\infty$ limit: the energy constraint and the evolution equation require that the spatial 
Ricci tensor is zero, see refs.~\cite{ellis71, mater} and ref.~\cite{buchertL} for a derivation in the tetrad formalism. This in turns implies that $\gg^\mathcal{L}_{\a\b}$ 
can be transformed to the Euclidean metric $\d_{AB}$ globally. In other words, at each time $\tau$ there exist global Eulerian observers comoving with the 
Hubble flow for which the components of the metric are $\d_{AB}$. This means that, according to the tensor transformation law, we can write the spatial metric as
\begin{equation} \label{globalflat}
 \overline{\gg}^\mathcal{L}_{\a\b}=\mathcal{J}^{A}_{\a}\mathcal{J}^{A}_{\b}\d_{AB}\,,
\end{equation}  
where $\mathcal{J}^{A}_{\a}$ is the Newtonian Jacobian matrix given by eq. \eqref{jacomap}. We can also find the transformation of the Christoffel symbols from the usual rule
\begin{equation}
\overline{\Gamma}^M_{LK}w^K_\mathcal{E}=\mathcal{J}^M_\nu\mathcal{J}^\rho_L\overline{\Gamma}^\nu_{\rho\sigma}w^\sigma_\mathcal{L}-\mathcal{J}^\rho_L \partial_\sigma\mathcal{J}^M_\rho w^\sigma_\mathcal{L}
\end{equation}
Since $\overline{\Gamma}^A_{BC}=0$ in Eulerian coordinates, we find the Christoffel symbols in Lagrangian coordinates 
\begin{equation}
\overline{\Gamma}^\a_{\b\sigma}=\mathcal{J}^\a_M \partial_\sigma\mathcal{J}^M_\b\,.
\end{equation}
We can therefore reformulate the Newtonian limit in this gauge, referring to the metric which results from the $c\rightarrow\infty$ limit of the Einstein equations: 
eq. \eqref{energy} and eq. \eqref{evolmater} tell us that we can write the spatial metric in the form of eq. \eqref{globalflat}. The Ricci tensor calculated from 
\eqref{globalflat} is zero but the Christoffel symbols involved in spatial covariant derivatives do not vanish. On the other hand, the vanishing of the spatial 
curvature implies that these covariant derivatives always commute. The resulting geometry in Lagrangian coordinates reproduces the properties of the Eulerian 
velocity and density fields, which come from the mapping \eqref{newtmap}: all it is needed is the Jacobian matrix, eq. \eqref{jacomap}, which is actually found by 
solving the remaining equations, the Raychaudhuri equation and the momentum constraint. They contain no explicit power of $c$, preserving their form in the Newtonian 
limit, and no curvature terms, which would involve higher (PN) terms of the metric. 

Now, we finally obtain the Newtonian expression of eq. \eqref{mom} and eq. \eqref{ray}, where by Newtonian expression here we mean the expression 
that comes from \eqref{globalflat}. Then, we rewrite the peculiar velocity-gradient tensor as 
\begin{equation} \label{thetanewt}
\overline{\vartheta}^\a_\b=\mathcal{J}^{\a}_{B}\mathcal{J}^{B'}_{\b}\,.
\end{equation}
The Raychaudhuri equation is therefore given by
\begin{equation}  \label{raynewt}
 \mathcal{J}^{\a}_{B}\mathcal{J}^{B''}_{\a}+\mathcal{H}\frac{\mathcal{J}'}{\mathcal{J}}=\frac{3}{2}\mathcal{H}^2\left(1-\frac{1}{\mathcal{J}}\right),
\end{equation}
where $\mathcal{J}$ is the determinant of the Jacobian matrix and $\left(1-\frac{1}{\mathcal{J}}\right)$ is the solution for the density contrast from 
eq. \eqref{deltasolution}.\footnote{We assumed for simplicity $\d_{in}=0$ and used the residual gauge freedom of the synchronous and 
comoving gauge to set $\mathcal{J}_{in}=1$ in the Newtonian limit, as in ref.~\cite{mater}.}
The momentum constraint reads
\begin{equation}
\overline{\mathcal{D}}_\a\left(\mathcal{J}^{\a}_{B}\mathcal{J}^{B'}_{\b}\right)=
\partial_\b\left(\frac{\mathcal{J}'}{\mathcal{J}}\right),
\end{equation}
where $\overline{\mathcal{D}}_\a$ is the covariant derivative related to the Newtonian metric, eq. \eqref{globalflat}.\\
On the other hand, eq. \eqref{thetanewt} together with $\overline{\vartheta}^\a_\b=(1/2)\, \overline{\gamma}^{\a\ss}\,\overline{\gamma}{\,'}_{\ss\b}$ gives
\begin{equation} \label{irronewt}
{{\cal J}_{A\alpha}}' {\cal J}^A_{~~\beta} = {\cal J}_{A\alpha} 
{{\cal J}^A_{~~\beta}}' 
\end{equation}
which is identical to the standard Newtonian form of the irrotational condition in Lagrangian space
\begin{equation}
\overline{\vartheta}_{[\a\b]}=0
\end{equation}
This equation, together with the relation $\partial_\b \mathcal{J}^A_\a=\partial_\a \mathcal{J}^A_\b$, which follows from the symmetry of 
the Newtonian Christoffel symbols, reduce the momentum constraint to an identity.

From the Newtonian limit of Einstein's equations in the synchronous and comoving gauge we then find eq.\eqref{raynewt} and eq.\eqref{irronewt}. 
These are identical to the well-known Lagrangian equations of Newtonian gravity, see e.g. ref.~\cite{catelan95L}, \cite{buchert89}, and~\cite{colombi92}.\\
A final remark about the energy constraint and the evolution equation: it would be wrong to take a Newtonian version of eq. \eqref{energy}, 
and eq. \eqref{evol} or eq. \eqref{evolmater}, by setting the l.h.s to zero. They simply imply that the Newtonian spatial curvature vanishes. On the contrary, 
they must be thought perturbatively: as a consequence of our gauge choice no odd powers of $c$ appear in the equations, so the expansion 
parameter is $1/c^2$. The spatial metric is then expanded up to PN order in the form
\begin{equation}
 \gg_{\a\b}=\overline{\gg}_{\a\b}+\frac{1}{c^{2}} w_{\a\b}
\end{equation}  
Therefore the Newtonian l.h.s of eq. \eqref{evol}, or of eq. \eqref{energy} and eq. \eqref{evolmater}, determine the spatial PN Ricci tensor, as shown in ref.~\cite{mater}.

\section{The transformation from the Eulerian to the Lagrangian frame}
In this section we provide the coordinate transformation for passing from the Newtonian limit in the Poisson gauge to the Newtonian limit in the synchronous and comoving gauge.

\subsection{The transformation of the spatial coordinates}

The Poisson gauge is defined in ref.~\cite{bert96} starting from the perturbed Einstein-de Sitter line element, the background spatial metric being $\delta_{ij}$. 
In comoving spatial Cartesian coordinates and conformal time the line element can be written in any gauge as 
\begin{equation}
ds^2=a^2(\eta)\left\{-\left(1+2\psi\right)c^2d\eta ^2 +2w_i cd \eta dx^i+\left[\left(1-2\phi\right)\delta_{ij}+2h_{ij}\right] dx^i dx^j\right\}
\end{equation}
where the tensor perturbation is trace-less and we have written explicitly the $c$ factor in the time coordinate. The four gauge modes are eliminating by setting
\begin{equation}
\begin{split}
\partial_i w^i=0  \,,
\qquad
\partial_i h^i_j = 0 \,,
\end{split}
\end{equation}
which fixes the Poisson gauge, including all the six physical degrees of freedom present in the metric. In particular, the Poisson gauge has no 
residual gauge ambiguity, since it can be shown that a coordinate transformation from an arbitrary gauge completely fixes this gauge. It is 
important to stress that in our approach all the degrees of freedom in the metric should be understood as {\it a priori} containing perturbations at any 
order in standard perturbation theory. The scalar potentials $\psi$, $\phi$ and the tensor $h_{ij}$ contain even powers of the speed of light, starting from $1/c^2$ and $1/c^4$, 
respectively. The vector $w_i$ contains odd powers of the speed of light, starting from $1/c^3$. We are free to change the time coordinate 
from $c\eta$ to $\eta$, since it just represents a change in the units of time, obtaining
\begin{equation}
ds^2=a^2(\eta)\left\{-\left(c^2+2\psi\right)d\eta ^2 +2w_i d \eta dx^i+\left[\left(1-2\phi\right)\delta_{ij}+2h_{ij}\right] dx^i dx^j\right\}
\end{equation}
The Newtonian limit in this gauge is obtained by retaining in the metric the only potential required in the Newtonian equations of motion, 
i.e. $\psi=\varphi_g$, as already explained in section~\ref{euler}. The Newtonian line-element in then
\begin{equation} 
ds^2=a^2\left[-\left(c^2+2\varphi_g\right) d\eta^2 +\delta_{ij}dx^i dx^j\right]\,
\end{equation}
For our purposes, it is useful to reinterpret this line element in the language of the $3+1$ splitting of space-time, see refs.~\cite{york} and~\cite{durrer3+1}, 
where the chosen coordinate system, i.e. the gauge, is related to the observers. In this formalism, the space-time is split in a family of three-dimensional 
hypersurfaces, the ``space'', plus the ``the time direction'', in strict analogy with the Newtonian treatment. On every three-dimensional hypersurface, 
the chosen time coordinate is constant, thus every hypersuface corresponds to the rest frame of the chosen observers. This perspective will of help here, 
since we are dealing with the Eulerian and Lagrangian frames. 

Eulerian observers are represented by a set of curves with unit four-velocity $n^a$ always orthogonal to the constant-time slices. 
Orthogonality implies that Eulerian observers are at rest on each slice and that there exists a scalar function $\mathcal{N}$, called the lapse function, such that
\begin{equation}
n_a=-\mathcal{N} \nabla_a\eta\;.
\end{equation}
It represent the rate of change of the proper time along $n^a$ with respect to the time $\eta$.
Given arbitrary three-dimensional coordinates on the initial slice, $\left\{q^\a\right\}$, one can construct a non-normal congruence threading 
the slices with tangent vectors $t^a$. Each curve is permanently labelled by the coordinate values it acquires on the initial slice. The vector 
basis related to the $\left\{q^\a\right\}$ is dragged along these curves, and not along the curves of Eulerian observers. The relation between the two 
four-velocities is given by
\begin{equation}
\begin{split}
t^a=\mathcal{N} n^a+\mathcal{N}^a\,, \qquad n^a\mathcal{N}_a=0 \;,
\end{split}
\end{equation}
where $\mathcal{N}^a$ is the projection of the velocity shift between the two frames on the slices. We can fix the shift vector such that 
$t^a$ coincides with the matter four-velocity $u^a$, namely we can choose the well-known comoving condition: in this case, a given element 
of the fluid has fixed spatial coordinates and the $\left\{q^\a\right\}$ are called Lagrangian. Of course, the time $\eta$ does not coincide with the 
proper (conformal) time defined in the rest frame of the fluid. The shift vector projected on the slices, i.e. on the rest frame of Eulerian observers, measures in 
$\eta$-time the spatial velocity of the matter with respect to the Eulerian observers. The line-element is that of the ADM formalism
\begin{equation}
ds^2 = a^2\left[-\mathcal{N}^2 d\eta^2 +g_{\a\b} \Big(dq^\a + \mathcal{N}^\a d\eta\Big) \left(dq^\b + \mathcal{N}^\b d\eta\right) \right].
\end{equation}
Instead of the basis related to the coordinates $\left\{q^\a\right\}$, we can alternatively choose an orthonormal basis on the spatial slices, 
whereas the time coordinate $\eta$ remains unchanged. The spatial coordinates are related in the usual way : $dx^A=\mathcal{J}^A_\a dq^\a$. The orthonormal 
basis is also dragged along the world-line of the corresponding fluid-element: the parallel transport condition of the tetrad $\mathcal{J}^A_\a$ along $u^a$ reads
\begin{equation} \label{paralleltrans}
\mathcal{J}^A_{\a _;a} u^a=0 \;,
\end{equation}
where the semicolon indicates the four-dimensional covariant derivative.
Of course, the orthonormal basis dragged along $u^a$ is not at rest with respect to the Eulerian observers: its relative velocity coincides with the peculiar velocity of the matter.
Using the orthonormal basis the line-element is
\begin{equation}
ds^2 = a^2\left[-\mathcal{N}^2 d\eta^2 +\delta_{AB}\Big(\mathcal{J}^A_\a dq^\a + \mathcal{N}^A d\eta\Big)\left(\mathcal{J}^B_\a dq^\b + \mathcal{N}^B d\eta\right)\right].
\end{equation}
In the Newtonian limit, the shift vector is the Newtonian peculiar velocity of the matter measured by Eulerian observers comoving with the Hubble flow, $\mathcal{N}^A=v^A$, 
and the lapse function is found to be $\mathcal{N}=1-2\varphi_g^\mathcal{E}$ from the time-time component of Einstein equations. 
The Newtonian line-element is then
\begin{equation}\label{ds2eul}
ds^2=a^2\left[-\Big(c^2+2\varphi_g^\mathcal{E}\Big) d\eta^2+\delta_{AB}\Big(\mathcal{J}^A_\a dq^\a + v^A d\eta\Big)\left(\mathcal{J}^B_\a dq^\b + v^B d\eta\right)\right]
\end{equation}
where the matrix $\mathcal{J}^A_\a $ is the Jacobian matrix of the map
\begin{equation}
\mathbf{x}(\mathbf{q},\eta)=\mathbf{q}+\mathbf{\mathcal{S}}(\mathbf{q},\eta)\,.
\end{equation}
The coordinates $x^A$ on the slices representing the rest frame of Eulerian observers are the usual Eulerian coordinates of Newtonian gravity. This is our starting 
point for the transformation to the synchronous and comoving gauge.
We transform from the Eulerian spatial coordinates $x^A$ to the Lagrangian ones $q^\a$ in the line-element \eqref{ds2eul}, obtaining
\begin{equation}\label{newtboostds^2}
ds^2=a^2\left[-\left(c^2+2\varphi_g^\mathcal{E}-v^Cv^D\delta_{CD}\right) d\eta^2 +2\delta_{AB}\mathcal{J}^A_\sigma v^Bdq^\sigma d\eta + 
\delta_{AB}\mathcal{J}^A_\sigma\mathcal{J}^B_\lambda dq^\sigma dq^\lambda\right].
\end{equation}
Note that at this step we have not changed the time coordinate, according to a purely Newtonian treatment where the time is absolute. The slices $\eta= \text{const.}$ still 
set the rest frame of the Eulerian observers, not yet the rest frame of the matter. On the $\eta= \text{const.}$ slices, we have performed a spatial transformation from the spatial 
orthonormal basis to the Lagrangian one, the two bases moving with relative velocity $v^A$ with respect to each other.
This transformation can be alternatively viewed as a boost with spatially varying relative velocity $v^A$ between the Lagrangian and the Eulerian frame: 
\begin{equation}
u^a=\Gamma\left(n^a+v^a\right)\,,
\end{equation} 
where $\Gamma=\left(1-v^2/c^2\right)^{-1/2}$ is the Lorentz factor. In the Newtonian limit, the unit normal $n^a$ has no spatial component and the coordinate three-velocity $v^A$ is that measured by the Eulerian observers in $\eta$ time and we have $\Gamma\simeq 1$. The boost is supplemented by a spatial coordinate transformation from the orthonormal coordinate basis related to the $x^A$ to that related to the $q^\a$. In matrix form, our resulting transformation reads
\begin{equation} \label{galileomatrix}
 \mathcal{A}^{a}_{b} = 
\begin{pmatrix}
 1&  &0 \cr
 & & & \cr
 v^A  && \mathcal{J}^A_\a\,.
 \end{pmatrix}
\end{equation}
Now, we need a second step to arrive at the synchronous and comoving gauge: we have to change from the Eulerian observers rest frame to the fluid rest frame. 

\subsection{The transformation of the time coordinate}
\label{transt}
We write the time transformation required to obtain the correct quantities in the Lagrangian frame as
\begin{equation}
\tau=\eta-\frac{1}{c^2}\xi^0_\mathcal{E}(x^A,\eta)
\end{equation}
and its inverse
\begin{equation}
\eta=\tau +\frac{1}{c^2}\xi^0_\mathcal{E}(x^A,\eta) \;,
\end{equation}
where we want the function $\xi^0_\mathcal{E}$ in terms of the $q^a$.
Note that the functions $\xi^0_\mathcal{E}$ and $\xi^0_\mathcal{L}$ are functions of the Eulerian and Lagrangian coordinates, respectively; 
the (conformal) time dependence at the required order is the same, $\eta=\tau$, and the spatial coordinates on the slices change as $dx^A=\mathcal{J}^A_\a dq^\a$. We have
\begin{equation}
\xi^0_\mathcal{E}(x^a(q^b))=\mathcal{J}^0_s\xi^s_\mathcal{L}(q^b)=\xi^0_\mathcal{L}(q^b)\,,
\end{equation}
thus $\xi^0$ transforms simply as a scalar under the change in spatial coordinates, i.e.
\begin{equation}
\xi^0_\mathcal{E}(x^A(q^\b),\eta)=\xi^0_\mathcal{L}(q^\a,\tau).
\end{equation}
Now, we go back to the line-element of eq. \eqref{newtboostds^2}. In order to get
\begin{equation}
ds^2=a^2\left[-c^2 d\tau^2 +\gamma^\mathcal{L}_{\a\b}dq^\a dq^\b\right]
\end{equation}
we substitute
\begin{equation}
\tau=\eta-\frac{1}{c^2}\xi^0_\mathcal{L}(\eta,q^\a)
\end{equation}
in $-c^2 a^2(\tau)d\tau^2$, obtaining
\begin{equation}
-c^2 a^2(\tau)d\tau^2=-c^2\left(a^2(\eta)-\frac{2}{c^2}a^2(\eta)\mathcal{H}\xi^0_\mathcal{L}\right)\left(d\eta-\frac{1}{c^2}d\xi^0_\mathcal{L}\right)^2 \;,
\end{equation}
which, at $1/c^2$ order reduces to
\begin{equation}
-c^2 a^2(\tau)d\tau^2=-c^2a^2(\eta)d\eta^2+a^2(\eta)\left(2\mathcal{H}\xi^0_\mathcal{L}+2\frac{\partial \xi^0_\mathcal{L}}{\partial\eta}\right)d\eta^2+2a^2(\eta)\frac{\partial \xi^0_\mathcal{L}}{\partial q^\sigma}dq^\sigma d\eta\,.
\end{equation}
Comparing with eq. \eqref{newtboostds^2}
\begin{equation}
ds^2=a^2(\eta)\left[-\left(c^2+2\varphi_g^\mathcal{L}-v^Cv^D\delta
_{CD}\right)d\eta^2+2\delta_{AB}\mathcal{J}^A_\sigma v^Bdq^\sigma d\eta+\mathcal{J}^A_\sigma\mathcal{J}^B_\rho\delta_{AB}dq^\sigma dq^\rho\right]
\end{equation}
we finally get the equations for $\xi^0_\mathcal{L}$
\begin{eqnarray}
2\mathcal{H}\xi^0_\mathcal{L}+2\frac{\partial \xi^0_\mathcal{L}}{\partial\eta}&=&-2\varphi_g+v^Av^B\delta_{AB} \label{eqxi0t}\\
v^{K}&=&\frac{\partial \xi^0_\mathcal{L}}{\partial q^\lambda}\mathcal{J}_F^\lambda\delta^{FK} \label{eqxi0k}\,.
\end{eqnarray}
Integrating eq. \eqref{eqxi0t} gives 
\begin{equation}
\xi^0_\mathcal{L}=\frac{1}{a}\int^\eta_{\eta_{in}}a\left(-\varphi_g^\mathcal{L}+\frac{1}{2}v^Av^B\delta_{AB}\right) d\tilde{\eta} +\frac{C(q^\a)}{a}\,,
\end{equation}
where $C(q^\a)$ is an integration constant which we will fix using \eqref{eqxi0k}: we re-write this equation as
\begin{equation}
\frac{\partial \xi^0_\mathcal{L}}{\partial q^\ss}=v^K\mathcal{J}^F_\ss\delta_{FK}\,, 
\end{equation}
then $C(q^\a)$ is found from
\begin{equation}
v^K\mathcal{J}^F_\ss\delta_{FK}-\frac{1}{a}\frac{\partial}{\partial q^\ss}\left[ \int^\eta_{\eta_{in}}a\left(-\varphi_g^\mathcal{L}+\frac{1}{2}v^Av^B\delta_{AB}\right) d\tilde{\eta}\right]=\frac{1}{a}\frac{\partial C}{\partial q^\ss}\,.
\end{equation}

We obtain the same expression for $\xi^0$ also following the procedure outlined in ref.~\cite{KMNR}, which exploits directly the fact that the time coordinate of the synchronous and comoving gauge 
is the proper time along the world-line of the fluid: in ref.~\cite{KMNR} the calculation was performed at second order in standard perturbation theory, whereas ours is not restricted to any perturbative order.
We will show in the next section that our solution for $\xi^0$, when expanded at second order, coincides with that of ref.~\cite{KMNR}.

\subsection{Four-dimensional gauge transformation of the metric}
The transformation between the Eulerian and Lagrangian frame is
\begin{equation} \label{galileomatrix}
 \mathcal{A}^{a}_{b} = 
\begin{pmatrix}
 1&  &0 \cr
 & & & \cr
 v^A  && \mathcal{J}^A_\a
 \end{pmatrix}\,,
\end{equation}
and the time transformation is
\begin{equation} \label{timematrix}
 \mathcal{B}^{a}_{b} = 
\begin{pmatrix}
 1+\dfrac{1}{c^2}\dfrac{\partial\xi^0}{\partial\tau} &  &\dfrac{1}{c^2} \dfrac{\partial\xi^0}{\partial q^\b} \cr
 & & & \cr
 0  && \delta^A_\a
 \end{pmatrix} \,.
\end{equation}
Their product gives the transformation of the space-time coordinates. The result is the four dimensional coordinate transformation: 
\begin{equation} \label{jacnewt}
 \mathcal{J}^{a}_{b} = 
\begin{pmatrix}
 1+\dfrac{1}{c^2}\dfrac{\partial\xi^0}{\partial\tau} &  &\dfrac{1}{c^2} \dfrac{\partial\xi^0}{\partial q^\b} \cr
 & & & \cr
 v^A  && \mathcal{J}^A_\b
 \end{pmatrix}\,.
\end{equation} 
We now calculate how the metric tensor transforms under the transformation $x^a\rightarrow q^{a}$: from the standard rule we have
\begin{equation}
g^\mathcal{L}_{ab}(q^{c})=\frac{\partial x^d}{\partial q^{a}}\frac{\partial x^f}{\partial q^{b}}g^\mathcal{E}_{df}(x^d(q^{c}))\,,
\end{equation} 
We also need the Newtonian Eulerian metric $g^\mathcal{E}_{ab}$
\begin{equation} \label{jacnewt}
 g^\mathcal{E}_{ab}= a^2
\begin{pmatrix}
 -\left(c^2+2\varphi_g^\mathcal{E}\right) &  0\cr
 & & & \cr
 0 & \delta_{AB}
 \end{pmatrix}\,.
\end{equation} 
Applying the synchronous and comoving gauge conditions to the time-time and space-time components on the l.h.s., i.e. $g^\mathcal{L}_{00}=-1$ and $g^\mathcal{L}_{0\a}=0$, 
and expanding up to $1/c^2$ we find the same equations for $\xi^0_\mathcal{L}$ that we found in section \ref{transt}: the equation for $g^\mathcal{L}_{00}$ is
\begin{equation}
2\mathcal{H}\xi^0_\mathcal{L}+2\frac{\partial \xi^0_\mathcal{L}}{\partial\eta}=-2\varphi_g^\mathcal{E}+v^Av^B\delta_{AB} 
\end{equation}
with solution
\begin{equation} \label{xi0solution}
\xi^0_\mathcal{L}=\frac{1}{a}\int^\eta_{\eta_{in}}a\left(-\varphi_g^\mathcal{L}+\frac{1}{2}v^Av^B\delta_{AB}\right) d\tilde{\eta} +\frac{C(q^\a)}{a}\,,
\end{equation}
where $C(q^\a)$ has to be fixed from the equation for $g^\mathcal{L}_{0\a}$
\begin{equation}
v^{K}=\frac{\partial \xi^0_\mathcal{L}}{\partial q^\lambda}\mathcal{J}_F^\lambda\delta^{FK}\label{dxxi0}\,.
\end{equation}
For the spatial components we get
\begin{equation} \label{newtgabL}
\overline{\gamma}^\mathcal{L}_{\a\b}=\mathcal{J}^A_\a\mathcal{J}^B_\b\delta_{AB}\,.
\end{equation}
The second equation for $\xi^0$ obtained from our transformation, eq. \eqref{eqxi0k}, is very important: it shows that $\xi^0_\mathcal{L}$ can be 
thought as the velocity potential in Lagrangian space\footnote{Remember that although the three-velocity vanishes in the Lagrangian space, the velocity-gradient tensor is well defined.}. 
From the irrotationality condition in Eulerian space we know that $v^A= \partial^A \Phi_v$ and the peculiar velocity-gradient tensor is $\vartheta^A_B= \partial^A\partial_B \Phi_v$. 
By changing the spatial derivative in Lagrangian coordinates according to $\partial_A=\mathcal{J}_A^\sigma\partial_\sigma$ and using 
eq. \eqref{eqxi0k}, it is easy to show that the peculiar velocity-gradient tensor in Eulerian space transforms to 
\begin{equation}
\overline{\vartheta}^\a_\b=\overline{\mathcal{D}}^\a\overline{\mathcal{D}}_\b\xi^0_\mathcal{L}
\end{equation}
in Lagrangian space. The Lagrangian velocity-gradient tensor appears as the spatial covariant derivative of a scalar, but when expanded at second order 
in perturbation theory it acquires a true tensorial part, owing the expression of the Christoffel symbol in Lagrangian space.

\section{Consistency with perturbation theory up to second order}

\subsection{Second-order Newtonian solutions in the Eulerian picture}

In this subsection we report the results for the dynamics of irrotational dust in the Eulerian picture. For the complete calculation, we refer to e.g. ref.~\cite{catelan95E}.
The second-order expressions for the peculiar gravitational potential, the peculiar velocity and the density contrast are
\begin{eqnarray}
\varphi_g^\mathcal{E}&=&\phi-\frac{5}{21}\eta^2\Psi_\mathcal{E}+\frac{\eta^2}{12}\partial^K\phi\partial_K\phi \label{varphigE}\\
\mathbf{v}_\mathcal{E}&=&\nabla\left(-\frac{\eta}{3}\phi-\frac{\eta^3}{36}\partial^K\phi\partial_K\phi+\frac{\eta^3}{21}\Psi_\mathcal{E}\right)\label{vE}\\
\delta_\mathcal{E}&=&\frac{\eta^2}{6}\partial^K\partial_K\phi+\frac{5}{252}\eta^4\left(\nabla_x\phi\right)^2+\frac{\eta^4}{126}\partial^K\partial^N\phi\partial_K\partial_N\phi+\frac{\eta^4}{36}\partial^K\phi\partial^K\nabla_x\phi\,, \label{deltaE} \;,
\end{eqnarray}
where $\phi$ is the peculiar gravitational potential evaluated at the initial time $\eta_{in}$ and the potential $\Psi_\mathcal{E}$ is given by
\begin{equation}
\nabla_x^{2}\Psi_\mathcal{E}=-\frac{1}{2}\left[\left(\nabla^2_x\phi\right)^2-\partial^N\partial^K\phi\partial_N\partial_K\phi\right].
\end{equation}
These expressions coincide with the time-time perturbation of the metric, the matter peculiar velocity and the density contrast in the Poisson gauge, at second order in GR perturbation theory, 
retaining the Newtonian terms only, see ref.~\cite{BMPR2010}.\\

\subsection{Newtonian transformed metric up to second order in the synchronous and comoving gauge}

The aim of this section is to calculate the spatial metric $g^\mathcal{L}_{\a\b}$ and the function of the time transformation $\xi^0_\mathcal{L}$, 
starting from the Eulerian field at second order in perturbation theory given by eqs. \eqref{varphigE}, \eqref{vE}, and \eqref{deltaE} in the last section.

For the second-order expansion we have to be careful: every term is a function $F_\mathcal{E}(x^a(q^b))$ which has to be expanded itself together with its argument. In order for  all the expressions to be Newtonian, 
all terms are functions of the coordinates $x^A=q^A+\mathcal{S}^A$, with absolute time $\eta=\tau$. Therefore, we have this two-step Taylor expansion: the second-order expansion of an Eulerian function is given by  
\begin{equation}
F_{\mathcal{E}}(x^a)=F^{(0)}_{\mathcal{E}}(x^a)+ F^{(1)}_{\mathcal{E}}(x^a)+\frac{1}{2}F^{(2)}_{\mathcal{E}}(x^a) \;,
\end{equation}
where we have to expand the argument with respect to the Newtonian perturbative transformation for the $x^a$:
\begin{equation}
\begin{split}
x^A =q^A +S^{A(1)}+\frac{1}{2}S^{A(2)}\,, \qquad \eta =\tau\,.
\end{split}
\end{equation}
By collecting all terms of the same order, we find the second-order expansion of the Lagrangian function. The result is
\begin{eqnarray} \label{Taylor2}
F_{\mathcal{L}}(q^a)&=&F^{(0)}_{\mathcal{E}}(q^a)+\left(\left.\frac{\partial F^{(0)}_{\mathcal{E}}}{\partial x^A}\right|_{x=q}S^{A(1)}+F^{(1)}_{\mathcal{E}}(q^a)\right)
+\frac{1}{2}\left(2\left.\frac{\partial F^{(1)}_{\mathcal{E}}}{\partial x^A}\right|_{x=q}S^{A(1)}+\right.\nonumber\\
&&\left.+ F^{(2)}_{\mathcal{E}}(q^a)+\left.\frac{\partial^2 F^{(0)}_{\mathcal{E}}}{\partial x^A\partial x^B}\right|_{x=q}S^{A(1)}S^{B(1)}+\left.\frac{\partial F^{(0)}_{\mathcal{E}}}{\partial x^A}\right|_{x=q}S^{A(2)}\right).
\end{eqnarray}
The FRW background, in particularly the Einstein-de Sitter background considered here, keeps its form in any of the two gauges considered
\begin{equation}
ds^2=a^2(\eta)\eta_{ab}dx^adx^b\,,
\end{equation}
where $\eta$ is the conformal time. The lapse function is simply the scale factor and the shift vector vanishes, so that the Eulerian and Lagrangian spatial coordinates coincide. 
In standard perturbation theory, the background spatial transformation is simply $dx^A=\delta^A_\a dq^\a$. So, in the following we can set the Latin indices equal to the Greek 
ones in the spatial derivatives of second-order quantities, whereas the spatial derivatives and the arguments of first-order quantities change with the Jacobian matrix of the first-order transformation.

We then follow an iterative procedure, starting from the Eulerian first-order peculiar velocity: the time integration gives the first-order spatial transformation
\begin{equation}
\mathbf{x} = \mathbf{q} -\frac{\tau^2}{6}\nabla_q \phi
\end{equation}
with Jacobian $\mathcal{J}^\a_\b=\delta^\a_\b-\tau^2/6 \,\partial^\a\partial_\b\phi$ and inverse $\mathcal{J}^\a_\b=\delta^\a_\b+\tau^2/6 \,\partial^\a\partial_\b\phi$, where $A=\a$ at first order.
Then we use the Eulerian velocity at second order to find the second-order spatial transformation: after changing coordinates in the first-order term and integrating over time we find
\begin{equation}
x^\a = q^\a-\frac{\tau^2}{6}\partial^\a\phi+\frac{\tau^4}{84}\partial^\a\Psi_\mathcal{L} \;.
\end{equation}

We can now find the solution for $\xi^0_\mathcal{L}$, eq. \eqref{xi0solution}: 
\begin{equation} 
\xi^0_\mathcal{L}=\frac{1}{a}\int^\eta_{\eta_{in}}a\left(-\varphi_g^\mathcal{L}+\frac{1}{2}v^Av^B\delta_{AB}\right) d\tilde{\eta} +\frac{C(q^\a)}{a}\,.
\end{equation}
In the integral, we have to transform the scalar $\varphi_g^\mathcal{E}$ with the Taylor expansion of the form of equation \eqref{Taylor2}. The result is 
\begin{equation}
\varphi_g^\mathcal{L}=\phi-\frac{10}{21}\tau^2\Psi_\mathcal{L}
\end{equation}
where now the potential $\Psi_\mathcal{L}$ is given by
\begin{equation}
\nabla_q^{2}\Psi_\mathcal{L}=-\frac{1}{2}\left[\left(\nabla^2_q\phi\right)^2-\partial^\sigma\partial^\lambda\phi\partial_\sigma\partial_\lambda\phi\right].
\end{equation}
The integration gives the solution $\xi^0_\mathcal{L}$ up to second order
\begin{equation}\label{xi02}
\xi^0_\mathcal{L}=-\frac{1}{3}\tau\phi+\frac{1}{21}\tau^3\Psi_\mathcal{L}+\frac{1}{36}\tau^3\partial^\sigma\phi\partial_\sigma\phi+\frac{C(q^\a)}{a}\,.
\end{equation}
We use the second-order Eulerian velocity and remaining equation obtained from our transformation, eq. \eqref{dxxi0}, to fix the constant $C(q^\a)$. It turns out to vanish at second-order, thus it is at least a third-order quantity.\\
Our final expressions
\begin{eqnarray}
\mathbf{x} &=& \mathbf{q}-\frac{\tau^2}{6}\nabla_q\phi+\frac{\tau^4}{84}\nabla_q\Psi_\mathcal{L} \label{spazio2}\\
\eta &=&\tau-\frac{1}{3}\tau\phi+\frac{\tau^3}{36}\partial^\sigma\phi\partial_\sigma\phi+\frac{\tau^3}{21}\Psi_\mathcal{L} \label{tempo2}
\end{eqnarray}
coincide with the second-order gauge transformation from the Poisson gauge to the synchronous and comoving gauge obtained from a fully relativistic calculation, 
see refs.~\cite{MMB}, \cite{rampf1} and \cite{rampf2}, retaining the Newtonian terms only. 
In particular, the time transformation also coincides with the result in ref.~\cite{KMNR}, where a different procedure was adopted.

Finally, our result for the Newtonian Lagrangian metric up to second order from eq. \eqref{newtgabL} and eq. \eqref{spazio2} is
\begin{equation}
g^\mathcal{L}_{\a\b}=\delta_{\a\b}-\frac{\tau^2}{3}\partial_\a\partial_\b\phi +\frac{\tau^4}{36}\partial_\sigma\partial_\a\phi\partial^\sigma\partial_\b\phi + 
\frac{\tau^4}{42}\partial_\a\partial_\b\Psi_\mathcal{L}\,.
\end{equation}
This is exactly the same expression obtained from the second-order solution of the Einstein equations in the synchronous and comoving gauge, considering only the Newtonian 
terms in the metric, see ref.~\cite{MMB}. It is important to note that this expression includes the contribution of second-order Newtonian tensor modes generated 
by scalar initial perturbations. It is well known that tensor modes appear already at the Newtonian and PN level in the metric in this gauge, whereas they are only PPN in the Poisson gauge. 

\section{Conclusions}
In this paper we examined the Newtonian approximation to the non-linear gravitational dynamics of cosmological perturbations.

Our starting point was the Newtonian line-element in the Poisson gauge, given by
\begin{equation}\label{gEconcl}
ds^2=a^2\left[-\left(1+2\frac{\varphi_g^\mathcal{E}}{c^2}\right)c^2 d\eta^2 +\delta_{AB}dx^A dx^B\right].
\end{equation}
We then transformed this metric to the Lagrangian frame and obtained the Newtonian line-element in the synchronous and comoving gauge as in ref.~\cite{mater}, which is given by
\begin{equation}\label{gLconcl}
ds^2=a^2\left[-c^2 d\tau^2 +\d_{AB}\mathcal{J}^{A}_{\a}\mathcal{J}^{B}_{\b} dq^\a dq^\b\right]\,.
\end{equation}
As we said, our starting point was the metric of eq.  \eqref{gEconcl} in the Poisson gauge, which we dubbed Newtonian, as the metric variables 
appearing there are just those needed for the Newtonian equations of motion. 
With our transformation we arrived at a Newtonian metric in the synchronous and comoving gauge, where, once again, we have just the variables needed for the Newtonian 
Lagrangian equations of motion, namely the spatial Jacobian matrix. However, the Newtonian three-dimensional space has vanishing spatial curvature, therefore, the Newtonian Lagrangian metric, 
eq. \eqref{gLconcl}, can be transformed globally to the Einstein-de Sitter background metric, 
the transformation being just the spatial transformation to Eulerian coordinates $dx^A=\mathcal{J}^a_\a dq^\a$, without changing the time coordinate. As we will see below, this inconsistency can be easily 
overcome by requiring that the Ricci four-dimensional curvature scalar is preserved by the transformation, as it should.

As we have shown, the Lagrangian frame can be transformed to the locally flat inertial frame by means of the transformation 
to coordinates $\left(\tau+1/c^2\,\xi^0, x^A\right)$:
\begin{equation} 
ds^2=a^2\left[-c^2 \left(d\tau+\frac{1}{c^2}d\xi^0\right)^2+\d_{AB}\mathcal{J}^{A}_{\a}\mathcal{J}^{B}_{\b} dq^\a dq^\b\right]\,.
\end{equation}
When our solution $\xi^0$, 
\begin{equation} \label{xi0solutionconcl}
\xi^0_\mathcal{L}=\frac{1}{a}\int^\eta_{\eta_{in}}a\left(-\varphi_g^\mathcal{L}+\frac{1}{2}v^Av^B\delta_{AB}\right) d\tilde{\eta} +\frac{C(q^\a)}{a}\,
\end{equation}
with $C(q^\a)$ fixed from
\begin{equation}
v^{K}=\frac{\partial \xi^0_\mathcal{L}}{\partial q^\lambda}\mathcal{J}_F^\lambda\delta^{FK}\label{dxxi0concl}\,,
\end{equation}
is expanded in $1/c^2$, at lowest order this line-element reproduces the Newtonian one in the Eulerian frame, eq. \eqref{gEconcl}.

On the other hand, the spatial scalar curvature $^{(3)}\!{\cal R}$ vanishes at lowest order in both frames. The conformal four-dimensional scalar curvature\footnote{We only give the conformal 
curvature here because, apart from the $a^{-2}$ factor, the extra term in the physical curvature is simply the Einstein-de Sitter scalar curvature in both gauges}, 
$^{(4)}\!{\cal R}\equiv\mathcal{R}$ of the metric \eqref{gEconcl} is given by
\begin{equation}
\mathcal{R}=-2\partial^A\partial_A\varphi_g^\mathcal{E}
\end{equation}
On the other hand, the conformal four-dimensional curvature in the synchronous and comoving gauge is given by
\begin{equation} \label{R4Dsynchro}
\mathcal{R}=2\vartheta'+\vartheta^2+\vartheta^\mu_\nu\vartheta^\nu_\mu+c^2\,^{(3)}\!{\cal R}
\end{equation}
and, at lowest order in our $1/c^2$ expansion, the PN spatial curvature contributes to the four-dimensional $\mathcal{R}$
\begin{equation}
\mathcal{R}=2\overline{\vartheta}'+\overline{\vartheta}^2+\overline{\vartheta}^\mu_\nu\overline{\vartheta}^\nu_\mu+\,^{(3)}\!{\cal R}^{PN}\,.
\end{equation}
In other words, at lowest order in our $1/c^2$ expansion, in the Eulerian frame, only the perturbation of the time-time component of the metric
contributes to the scalar curvature $\mathcal{R}$, whereas in the Lagrangian frame we need also the spatial PN term coming from the spatial PN metric: setting $\,^{(3)}\!\mathcal{R}=0$ in eq. \eqref{R4Dsynchro} at the lowest order would be incorrect. This very fact is clear from the PN expansion which actually shows that the metric contributes at different orders to the four-dimensional curvature.\\ 
In order to find the correct four-dimensional curvature in the Lagrangian frame, one cannot use the metric in eq. \eqref{gLconcl}, since the required PN part is missing.
This means that, in order to obtain the same expression for the scalar curvature $\mathcal{R}$, at lowest order after the change of frame,
we need to start from the weak-field metric in the Poisson gauge:
\begin{equation}
ds^2=a^2\left[-\left(1+2\frac{\varphi_g^\mathcal{E}}{c^2}\right)c^2 d\eta^2 +\left(1-2\frac{\varphi_g^\mathcal{E}}{c^2}\right)\delta_{AB}dx^A dx^B\right]\,,
\end{equation}
where only the scalar PN mode is considered in the metric, since vector and tensor modes give higher-order contributions to $\mathcal{R}$, once transformed to the Lagrangian frame. 
The transformation to the Lagrangian frame finally leads to
\begin{equation}
ds^2=a^2\left[-c^2 d\tau^2 +\left(1+\frac{\chi}{c^2}\right)\mathcal{J}^{A}_{\a}\mathcal{J}^{B}_{\b}\d_{AB}dq^\a dq^\b\right],
\end{equation}
where, see ref.~\cite{mater},
\begin{equation}
\chi= 2\mathcal{H}\xi^0_\mathcal{L} -2\varphi_g^\mathcal{L}-\Upsilon^\mathcal{L}.
\end{equation}

In the latter expression, the potential $\Upsilon^\mathcal{L}$ is given by
\begin{equation}
\overline{\mathcal{D}}^\ss\overline{\mathcal{D}}_\ss\Upsilon^\mathcal{L}=-\frac{1}{2}\left(\overline{\vartheta}^2-\overline{\vartheta}^\mu_\nu\overline{\vartheta}_\mu^\nu\right),
\end{equation}
and the peculiar velocity-gradient tensor is written in terms of our solution $\xi^0_\mathcal{L}$, eq. \eqref{xi0solutionconcl}, as 
$\overline{\vartheta}^\mu_\nu=\overline{\mathcal{D}}^\mu\overline{\mathcal{D}}_\nu\xi^0_\mathcal{L}$. 
The PN scalar mode $\chi$ comes from the transformation of the time coordinate, keeping only the scalar contributions in the PN spatial metric.

Following this procedure, starting from the scalar curvature $\mathcal{R}=-2\partial^A\partial_A\varphi_g^\mathcal{E}$ in the Eulerian frame 
we arrive at the same expression $\mathcal{R}=-2\overline{\mathcal{D}}^\a\overline{\mathcal{D}}_\a\varphi_g^\mathcal{L}$ in the Lagrangian frame.

Let us now compare our approach with that of ref.~\cite{MPS1993} and ref.~\cite{MPS1994}. Both papers deal with relativistic dynamics and  
consider the parallel transport condition, eq. \eqref{paralleltrans}, which in the Lagrangian approach of the synchronous and comoving gauge becomes
\begin{equation}
\mathcal{J}^{A'}_\a=\vartheta^\ss_\a\mathcal{J}^{A}_\ss\,.
\end{equation}
Then they solve the Einstein equations and find the velocity-gradient tensor\footnote{In ref.~\cite{MPS1993} the Einstein equations are integrated numerically 
in the special case of the so-called "silent Universe", see refs. \cite{MPSprl94} and \cite{BMP95}, and solved analytically for the plane-parallel dynamics. 
In ref.~\cite{MPS1994} the calculation is performed at second order in relativistic perturbation theory.} and the Jacobian matrix. 
Finally the Eulerian trajectories and peculiar velocity are obtained from
\begin{equation}
dx^A= \mathcal{J}^{A}_\ss dq^\ss\,.
\end{equation}
In this paper instead we have reconstructed the Lagrangian dynamics from the Eulerian fields in the Newtonian limit, 
thus our procedure actually goes in the opposite direction.

To conclude, let us emphasise once again that our transformation is different from the standard perturbation theory one.
In standard perturbation theory, the background spatial transformation is simply $dx^A=\delta^A_\a dq^\a$;
in a relativistic calculation, the same background transformation from the Eulerian to Lagrangian frame is simply the standard Lorentz transformation, 
with boost velocity equal to the first-order peculiar velocity, see ref.~\cite{bert&hamilton}.
Instead, in the Newtonian limit, we have performed a non-trivial spatial transformation from the Eulerian to the Lagrangian frame -- the two 
frames moving with relative velocity $v^A$ -- and we have subsequently 
changed the parametrisation of the hyper-surfaces from the rest frame of Eulerian observers to the rest frame of the matter. 
This transformation is fully non-perturbative and will provide the background transformation for the same procedure in the PN approach.

A different approach to the Newtonian limit, see ref.~\cite{brunidrag}, is to consider the lower order in the $1/c^n$ expansion of all 
the Einstein equations, thus including in the Newtonian metric the scalar $\mathcal{O}(1/c^2)$ in the spatial part and the vector 
$\mathcal{O}(1/c^3)$ in the time-space component in the Poisson gauge. The difference between these two approaches is just semantic: 
in the Poisson gauge, what we call PN in our viewpoint is the same as what is called Newtonian in ref.~\cite{brunidrag}. The present transformation has to 
be considered the background transformation for the PN expansion.
Nevertheless, it clarifies the meaning of the change of space and time coordinates from the Eulerian to the Lagrangian frame.

\acknowledgments
We thank Julian Adamek, Daniele Bertacca, Marco Bruni and Cornelius Rampf for helpful discussions. Eleonora Villa thanks "Fondazione Angelo della Riccia" for financial support.

\end{document}